# The Lomonossov arc: refraction and scattering in Venus atmosphere during solar transits

by

Serge Koutchmy

*Institut d'Astrophysique de Paris (IAP)- CNRS and Sorbonne Université*

*98 Bis Boulevard Arago Paris F-75014 (France)*

koutchmy@iap.fr

**Abstract:** The main observations of 1761 by M. Lomonossov and those that followed are recalled by extending the discussion to other remarkable visual observations of the passages, then with more and more powerful imagers producing images in profusion. The modern treatment of parasitic effects is briefly recalled by focusing on the expert observation of 1761 which has recently been widely commented on and criticized. It included a spurious effect called the "black drop effect". The shell or aureole or atmospheric ring of Venus observed outside the solar disk is considered with reference to the today parameters of the Venus atmosphere. The contacts during the transit are discussed taking into account effects of scattering, absorption and the dominant effects of the refraction at the small angular distances found to be comparable to a fraction of the angular dimension of the planet. Modern observations of the 2004 and the 2012 transit are tentatively discussed to elucidate what is the arc of Lomonossov.

**Keywords:** Venus atmosphere; transit; refraction effect; arc; aureola; aerosols; Lomonossov ; extra-solar planets; habitable worlds

# Outline:

## I- Introduction

## II- Some historical aspects on the evaluation of Lomonossov's role.

II-1- Lomonossov, an outstanding figure

II-2- In 1761, the challenge of an unsuccessful method

II-3- Lomonossov: The Astronomer

## III- Observation of the refraction effects during the passages of Venus in front of the Sun.

III- 1- Lomonossov's observations of 1761

III-2- Other observations from 1761 to 1882

III- 3- Modern observations

III- 4- The black drop parasitic effect.

## IV- The atmosphere of Venus confirmed

IV-1- Confirmation after 1761

IV- 2- The Venus atmosphere today.

## V- On the modeling of the refraction phenomenon in the atmosphere of Venus and the important associated effects.

V- 1- The classical treatment: a transparent atmosphere

V- 2- Some simple naive considerations to start with.

V- 3- An atmosphere with scattering and absorption

V- 4- An improvement of the model.

## VI- Modern measurements of the Lomonossov arc.

VI- I- The ground-based high-resolution observation of 2004

VI-2- The space observation of 2012

## VII- Conclusion

## Acknowledgements

## References

# I- Introduction

During the transit (or passage) of the planet Venus in front of the Sun on June 6, 1761, Academician M. Lomonossov from St Petersburg announced the evidence of a thick atmosphere around the planet (<1><2> <4>). This interpretation turned out later to be correct: solar rays reaching the observer are significantly deviated by refraction around the planet during its passage at the solar limb, as viewed in projection on the sky. Lomonossov's interpretation is far from obvious, especially for the time, and the announcement must be placed in its historical context, since passages deserved study as they had important cosmogonic consequences.

The modern treatment of these effects, which are certainly more easily observed in the context of lunar eclipses, is briefly recalled by focusing on the expert observation of 1761. This observation has recently been widely commented and criticized, as it included a spurious effect, called the "black drop effect" and produced by the lack of angular resolution when imaging a sharp edge, as also observed during passages of the planet Mercury. The shell (or aureole or atmospheric ring) of Venus is also observed when the planet is viewed outside the solar disk, i.e. for rays coming from the Sun and deviated in both directions towards the observer. Another phenomenon - the forward Mie scattering by the aerosols of the upper atmosphere of Venus - can also be observed during its lower conjunctions. At the moment of contacts during the transit, these scattering effects can possibly be added to the dominant effects of the refraction, at angular distances remaining small compared to the planetary diameter. Finally, the adverse effect of absorption, especially in the infrared, may have to be accounted for.

A quite similar situation is met when interpreting the light curve of the transit of a giant exoplanet across the disk of a distant star (Hubbard et al <5>), but in case of Venus, the planet is small and rather distant from the parent star.

Given the current and important searches for signature of atmospheres during the transits of exoplanets, habitable or not, in front of their parent star, it reinforces the need to understand the various signatures observed at the occasions of the very rare transits of Venus in front of the Sun, as seen from Earth or low orbit satellites.

## II- Some historical aspects on the evaluation of Lomonossov's role.

There is an abundant literature relating and discussing what is nowadays called "the Lomonossov effect" <1><2> or "the Venus halo" <3> or, better and more precisely, the Lomonossov arc <4><6>, these references being considered as fairly representative. Yet, it seems unthinkable to analyze the discovery of an atmosphere around a planet other than our Earth, without placing it first in the context of the time, more than two and a half centuries ago.[1]

### II-1- Lomonossov, an outstanding figure

The French Wikipedia article dedicated to Mikhail Lomonossov (ML) see [Mikhail Lomonossov - Wikipedia (wikipedia.org)](#) is a very good and easily accessible source to get to know the character. The article is largely based on the quite fundamental and somewhat committed work of Luce Langevin <6> who paints a laudatory and edifying portrait of the man without, however, dwelling on his achievements in optics nor, moreover, on his observation of Venus transit in front of the Sun in 1761. The article in English on Wikipedia see [Mikhail Lomonosov - Wikipedia](#) is more explicit with a chapter on the physicist, then on the astronomer, with surprisingly no reference to the French work of Luce Langevin. Comparing these two considerable texts of Wikipedia already shows the ambiguities on the evaluation of Lomonossov's role, without speaking about the article of Wikipedia in Russian which is still different and certainly more "Lomonossovian" see [Ломоносов, Михаил Васильевич — Википедия (wikipedia.org)](#) . It would take many pages to discuss the matter in depth. In short, we can arbitrarily but probably objectively quote these few lines taken from one of the most popular "English Wikipedia" articles:

---

[1] Note that the spelling of the name has been adapted in English <4> <8>, which made it lose an "s" but it is the phonetically more correct French spelling that has been kept in this article or often the whole name will be replaced by "ML", to avoid too many unnecessary repetitions.

*"Mikhail (Mikhaylo) Vasilyevich Lomonosov ([/ˌlɒməˈnɒsɒf/](#);[\[1\]](#) [Russian](#): Михаил (Михайло) Васильевич Ломоносов; November 19 [[O.S.](#) November 8] 1711 - April 15 [[O.S.](#) April 4] 1765) was a [Russian polymath](#), scientist and writer, who made important contributions to literature, education, and science. Among his discoveries were the [atmosphere of Venus](#) and the law of [conservation of mass](#) in [chemical reactions](#). His spheres of science were [natural science](#), [chemistry](#), [physics](#), [mineralogy](#), history, art, [philology](#), [optical devices](#) and others. Founder of modern [geology](#),[\[2\]](#)[\[3\]](#) Lomonosov was also a poet and influenced the formation of the modern [Russian literary language](#). "*

It is thus possible to classify him as one of the great encyclopedists of the 18th century whose influence was very significant during his lifetime in his country, as for the reform of the Russian language and the creation of the Moscow University bearing his name and which today includes an important Institute conducting research in astronomy and astrophysics, (figure 1). Besides, the article in French of Wikipedia notes :

*"A convinced patriot, an enthusiastic [polymath](#), a lover of science, he was a professor at the [Academy of Sciences of St. Petersburg](#) and founder of the [University of Moscow](#) (which bears his name); [Alexander Pushkin](#) even said of him that he was "the first university in Russia"* (author's translation).

When reading Lomonossov's report, written practically on the spot, of the observation of the arc phenomenon at the time of the contacts during the transit of 1761, it is not out of place to remember him as a poet and brilliant discoverer.

**II-2- In 1761, the challenge of an unsuccessful method**

For most of the many astronomers who observed the Venus transit of 1761, the main goal was to carry out a precise dating of the different contacts, in order to determine with parallax measurements taking advantage of observations made from sites sufficiently separated in longitude, the absolute distances between the celestial bodies Earth-Venus-Sun. At that time, this was felt of fundamental cosmogonic importance. One of the recommended methods had been discussed and proposed by the great E. Halley, who died in 1742 - thus before the transit of Venus. Halley had a great influence during his lifetime and his intellectual heritage was widely propagated during the whole 18th century. In France, J-N. Delisle was a fervent supporter to observe transits of Mercury and even more so, of Venus <7>.

These measurements of contacts in 1761 and thereafter have been the subject of countless interpretations as they are at least affected by significant errors with respect to the sought precision. The fault is possibly in the existence of this thick atmosphere of Venus which completely hides the planet's ground and makes questionable to use for Venus the method proposed by Halley, while it worked for the frequently observed transits of Mercury, this planet having no atmosphere. When applied to the rare couples of transits of Venus, these dating methods finally proved to be less precise than other astrometric methods applied to the same problem. Yet, one will not blame the illustrious promoter who so brilliantly succeeded on the comet bearing his name!

Venus transits remain so spectacular and rare that these measurements have lasted for at least the two couples of transits following those of 1761 and 1769, leading to the famous if not considerable expeditions in 1874 and 1882, and even nowadays in 2004 and 2012. Let simply observe that the ML discovery of a "rather thick" or dense atmosphere around Venus could have led to reconsider these expensive and often strictly astrometric efforts for distant missions.

**II-3- Lomonossov: the Astronomer**

During the years preceding the famous Venus transit of 1761, and even after, ML had worked on the construction of a telescope intended for twilight and night observation onboard ships, as well as on an off-axis telescope with a rather ingenious focal occulting system. There he was the forerunner of what is today named the Herschel type telescope, possibly not rendering justice to ML. This project was partly intended to observe the next transit of 1769, which he could not carry as he died in 1765. Being philosophically convinced of the plurality of the Worlds, he was opposed to the Russian Church and its hierarchy, at a time when the Russian translation of Fontenelle's book had been censored in Russia (1756). This and other subjects earned him academic animosities which hindered his activity. He was certainly convinced of the existence of an atmosphere around Venus, which he considered as the sister planet of the Earth. Therefore, the transit of Venus was for him a well-prepared observation, essentially to demonstrate the existence of such atmosphere through imagery.

To conclude this historical overview of the discovery of the Venusian atmosphere from the observation of transits, it is necessary to recall the strong links existing then between the Academy of Saint Petersburg and other Academies, especially with the French Academies. Voltaire was an associate member of the Russian Academy during approximately 32 years;

the famous astronomer J-N. Delisle worked for several years in St. Petersburg, where he had a great influence until 1745, when he had academic problems and returned to France. Later, he favored the long -over 2 years- and famous mission of Abbé Chappe d'Auteroche, immediately after his election in 1759 to the Academy of sciences, succeeding to Lalande [Wikipedia and Chappe (cosmovisions.com)]. This mission led Chappe to Tobolsk in Siberia, i.e. well beyond the Urals, to observe with great success the Venus transit of 1761. Chappe, very prolific in other respects, is nevertheless famous for reporting facts that he himself did not always live.

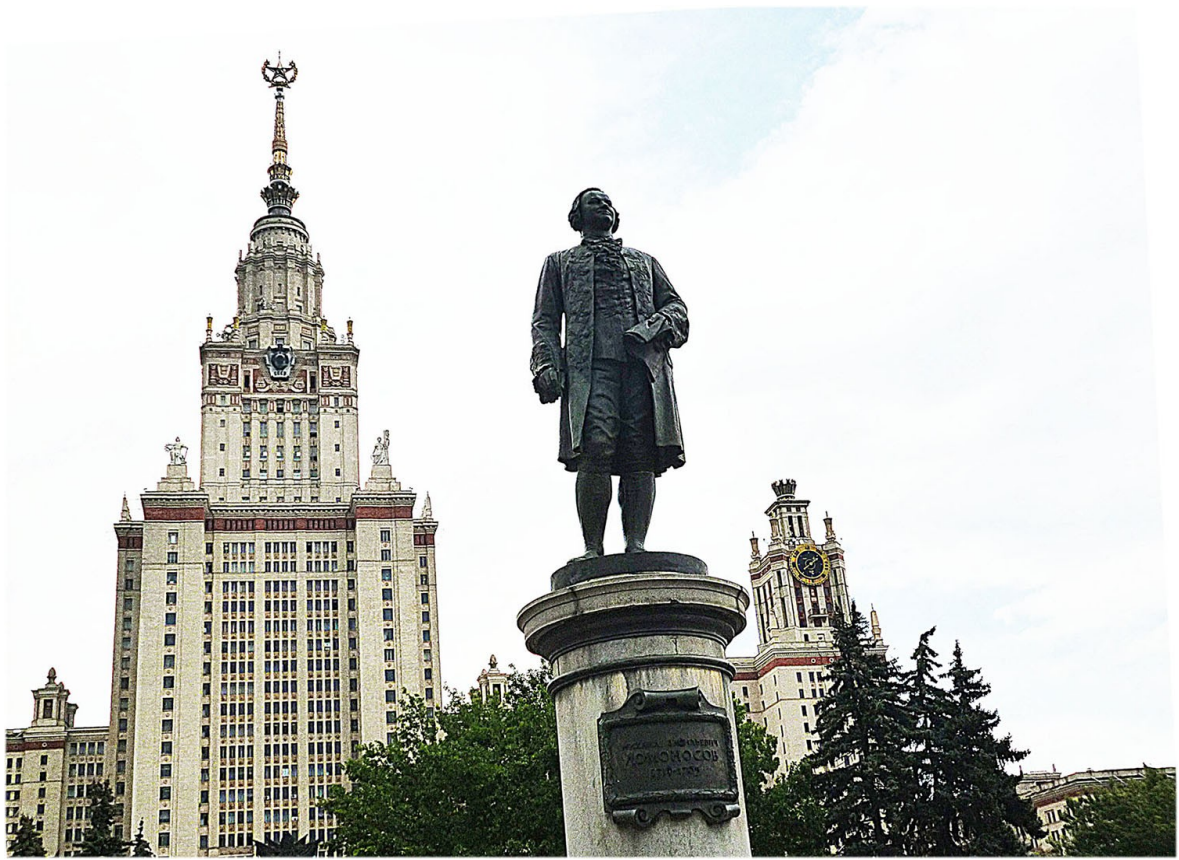

**Figure 1-** *Statue of Lomonossov erected in front of the Moscow University that he created in the 18th century (Photo taken by the author on the occasion of the 2014 COSPAR congress). ML is often considered in Russia as the first real scientist of this country.*

Finally, skipping almost two centuries let mention that, on the occasion of the bi-centenary of the announcement of the discovery of the Venus atmosphere by ML, the USSR launched towards Venus its first heavy satellite of the series "Venera", intended to analyze this atmosphere. In 1975, Venera 9 transmitted the first and still unique optical images of the planet's ground. It is likely that Lomonossov work contributed setting the priority given by USSR to its enormous space program towards Venus, perhaps to the detriment of the study of the planet Mars by this country.

## III- Observations of the refraction effects during the passages of Venus in front of the Sun.

**III- 1- Lomonossov's observations of 1761**

It is after observing the transit of Venus on June 6, 1761, that ML announces to have for the first-time observed effects due to refraction in the atmosphere of a planet passing in front of the Sun and, consequently, that the planet possesses an atmosphere. This was a crucial point considering the state of knowledge at the time. Let notice that he describes these effects only in connection with the observation of the contacts[2] and in particular just after the third contact, lasting about 1 minute, reporting an effect that he himself calls « pupyr » (as transcripted from Russian), a word that can be translated as "bump" or "blister" or "bulge", protruding at the edge of the Sun, as seen on his Figs. 3 and 4 (figure 2).

A detailed description of what he saw has been published several times: e.g. the most serious and recent paper by V. Shiltsev <4> containing many references to and discussion of previous papers; <9> comments the original ML paper, making this very short and critical statement : « ...*Mr. Councilor Lomonosov concludes that the planet Venus is surrounded by a significant atmosphere of air similar to that which surrounds our terrestrial globe.* ». Figs.1 to 8 in figure 2 appear to speak enough by themselves. Let underline the famous Fig. 7 giving the basic scheme of the refraction effect that ML has been the first to propose in the case of the atmosphere of a planet different from Earth, including the bending of rays coming from the Sun.

---

2  Contacts observed during the transit of a planet before the solar disk are supposed to correspond to moments i/ when the planet seems first to touch the Sun (1st contact) also called the ingress; ii/ when after, the planet get fully immersed (2d contact); at the exit of the planet, also called the egress, there are, well after, the symmetric 3d and 4th contacts.

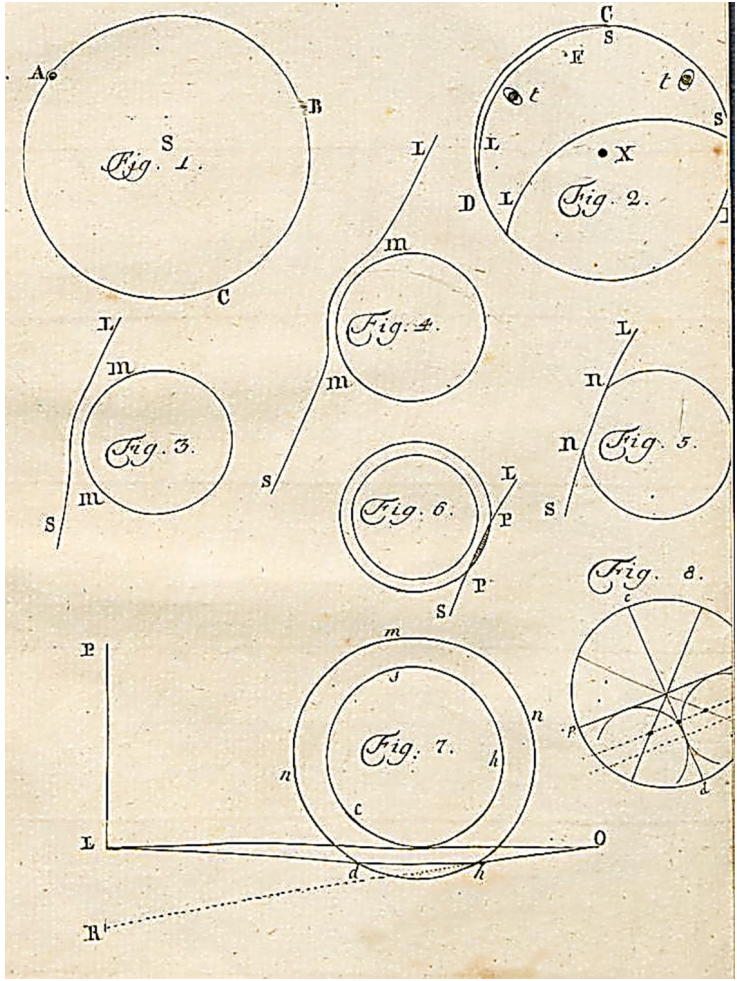

**Figure 2-** *Reproduction of the plate of the original drawings of ML provided in his report written after the observation of 1761, published in Russian and German <9> after 3 months. Visual observation made with one of the first refractors of Dollond (England) with a smoked glass of low optical density held at the focus <4>. The "Fig. 7" of the plate illustrates the interpretation given by ML in terms of refraction of the light rays coming from the Sun through the rather thick atmosphere supposed to exist around Venus and represented by an extended shell. It is the basic schematic of the refraction effect as studied after in many papers, using the bending of rays by refraction. No "black-drop" effect reported.*

**III-2- Other observations from 1761 to 1882**

In fact, about sixty expeditions <1><2> and more than 110 other observation sites can be counted in 1761 on the occasion of this rare astronomical event (period of 113 years) which takes place on different parts of the globe twice with an eight years interval. In 1761 the passages ~~are~~ were not visible from France, but Eastern Europe and Siberia offered many possibilities. A review of selected observations has been made by several authors, for example Fr. Link <1><2> <7> in his well-known article devoted to an interpretation of the phenomenon, or more recently by V. Shiltsev in his extensive article focusing on the 1761 passage <4>. As each observer has his own perception of the visually observed phenomenon, one might as well say that the testimonies diverge. Afterwards it is tempting to make a selection and only keep what seems plausible, but this is moving away from a strictly scientific method. Thus, many authors reporting the observations of 1761, 1769 then 1874 and 1882, have described quite systematically to see a halo or ring around the planet in projection on the solar disk. This appears doubtful for today's knowledge (see section VI below). These claims are presumably to be rejected due to their limited spatial resolution.

Nevertheless, surprisingly, let consider the arc appearing at the second contact, i.e. during the ingress or immersion when the planet is almost completely superposed to the solar disk, or at the third contact, i.e. during the egress or exit when the planet starts to leave the solar disk. This arc is rather well reported in 1761, but not so well in 1769, perhaps contrary to logic. In 1874 and later, the arc and the thin envelope or aureole around the planet seen after the first contact and before the fourth are perfectly described in a convincing way by Lagrange et al. <10> observing in almost perfect conditions in Chile. Their observations thus definitively confirmed the discovery of ML. Nevertheless, as noticed by Fr. Link <1><2>, ML's drawings of the arc are not without ambiguities and they leave an open door to criticism <1><2> <4> <12>.

Chappe d'Auteroche, already mentioned above for his fascinating trip to Siberia which lasted more than a year and his observation of the 1761 transit, reports later in 1762-1768 <7> <10> what seems to be beautiful observations of the phenomenon of aureole or ring and perhaps even an arc. He used two remarkable refractors of different focal lengths, equipped with a micrometer.

In his successive reports to the Academy of sciences, he discusses at length how his measurements of different Venus diameters at ingress and egress, compared to the values reported by other observers. He also discusses the effects appearing around the disk of Venus in projection on the solar disk, which appear rather confused when related to the modern observations: e.g. he draws a ring or aureola around the planet when well projected on the disk. As he was focusing on measurements with his web-wire micrometer, it is possible that he was a victim of a Mach effect induced by some defocusing (for a description of the Mach effect, see[3]). At the transit of 1769 he did not see this halo phenomenon at all and reports instead a "black drop" effect (see below) during the second and third contacts (ingress and egress), which does not fail to raise question [Chappe. (cosmovisions.com)](Chappe. (cosmovisions.com)). He never noted the role of refraction in a possible planetary atmosphere.

Unfortunately, he died shortly after the transit and his descriptions of 1761-62 were subsequently taken up again but in second hand.

The observation of the transit of 1882 by the Lagrange group (Royal Observatory of Brussels in Uccle) deserves a special mention because it includes a visual observation relation <11> (pp. 118 to 120 of the report) made with an achromatic refractor called Dollond's of the same nature as the one used by ML in 1761 and with the same method. The Dollond's refractor was invented and then produced in England a few years earlier. The author is the leader of the Belgian mission; the chosen site in Chile, near Santiago's Observatory, is excellent for astronomical observation, often having quiet and sharp images as claimed by Lagrange et al.

The instrument is of presumably of higher quality than the one used in 1761 by ML but the method is quite comparable, for example with the use of a blackened (smoked) glass near the focus. No mention of refraction effects is made by the authors of 1882. With a better sky than that of Saint Petersburg and without almost any image motions according to their report, the details reported on the shell or aureole and on the arc (in fact called 'horns') are certainly remarkable, not so far of what is done much later in 2004 and 2012 with large telescopes (see below). This is the case for the thickness of the envelope or aureole considered as very thin and stable by these authors; they call "light net" (namely "filet lumineux" in the original report in French). The plates of images attached to the report (e.g. Figure 3) are edifying if one puts oneself in the context of the description of ML of 1761 see <4> <9>.

---

[3] The Mach effect is an optical illusion, due to retinal lateral inhibition, which causes edges of darker objects adjacent to lighter objects to appear lighter than they truly are and vice versa thereby creating a false shadow, which may mimic injury or disease, etc. (from Wikipedia  Mach bands - Wikipedia  ).

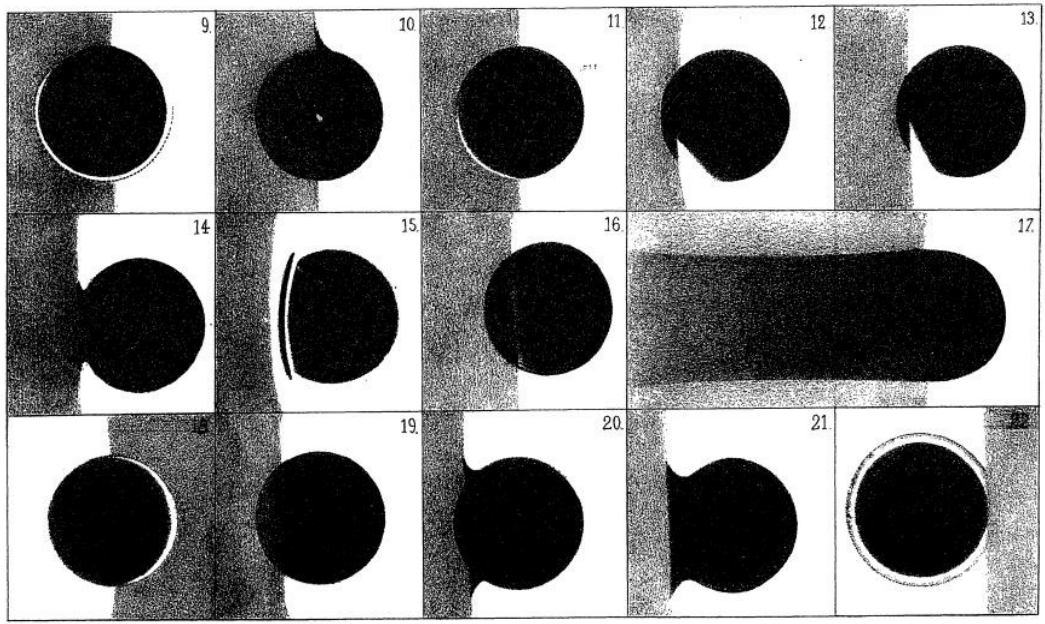

**Figure 3-** *Plate extracted from the report of the mission to Chile of the Brussels Observatory <11> during the transit of 1882. Visual observations made with excellent instruments, like Dollond's achromatic refractor of about 1 m focal length, in a clear sky with visually stable images, at least during the ingress. Images 9 to 16, during the ingress part of the transit; image 18, during the egress. Image 9 seems to have exaggerated the perception of the envelope or aureole, especially in projection on the solar disk and as an extension of the arc seen outside the Sun. Image 18 of the arc on 1/2 of the planet's circumference seems to be an erroneous a disputable representation of reality because the arc due to refraction is always is presumably thinner (indeed "less intense") at the top due to the limb darkening at the limb of the Sun.*

This point deserves some careful comment. Focusing on images 9, 11, 15 and 18 the assertions of ML's article (see <4> <9>) become clearer and justified; while, as shown in drawing 22 of this plate, the entire aureole or envelope is shown in projection on the disk, this seems hardly credible and possibly due to the Mach effect. The "black drop" effect is illustrated on pictures 20 and 21 taken at the end of the observations when the seeing became definitely poor; we discuss this below because this effect has nothing to do with refraction. Finally, image 15 also recalls one of the descriptions of ML at the time of the second contact of 1761 and it confirms, with drawings 12 and 13, the existence of effects which are difficult to interpret and sometimes at the origin of confusions, as described by Link in <1><2>, on the evaluation of the set of observations. Let us however notice that the interesting image 15 looks like the superposition of the black-drop effect and the effect of the arc seen at the extreme limb.

**III- 3- Modern observations**

To better understand the effect of refraction during a solar transit of Venus during the crucial arc phase, it is now interesting to extract images from contemporary observations made from space. Indeed, they are free of the parasitic effects produced by the Earth's atmosphere, turbulence and stray light. During the 2004 transit, the NASA TRACE mission telescope observed in several spectral channels, including a so-called "visible" channel. Its images were widely used by Pasachoff et al. (in 2004 see <12> and especially in 2011, see <13>), who also gives a good description of the TRACE mission imaging instrument, as well as important parameters for image evaluation. The pixel size gives an excellent resolution, better than 1" on the disk images, while not so good or more difficult to evaluate on the arc or aureole where the signal to noise is much worse. Unfortunately, only the qualitative aspects of the observation of transits are really discussed by Pasachoff and Schneider. More annoying, only composite images are given, as a disk fraction "juxtaposed" to the fraction outside the disk and amplified to appear with the same intensity. This is obviously far from reality see <6>. The discussion is limited to these images which, although excellent and "speaking", do not allow an objective comparison with the visual data of historical transits, including that of 1761. Moreover, the images presented by Pasachoff et al <13> are "cleaned" of instrumental effects such as ghost images due to spurious reflections of instrumental origin.

For these reasons, in Figure 4 an unretouched image from the originals is shown with amplified contrast to indicate what is perceived near a contact (second or third). Here the arc is shown by overexposing the disk and the result may be comparable to what ML and other historical observers described as a protruding arc from the solar limb. Using such image, it is possible to guess what is perceived when the resolution is less good, as for the instruments used in 1761 and after, and the influence of the atmosphere more significant, a factor difficult to apprehend. After some tests, it seems that an angular resolution of about 3" to 4" is sufficient to justify the historical assertions, provided that the sky is clear without any clouds.

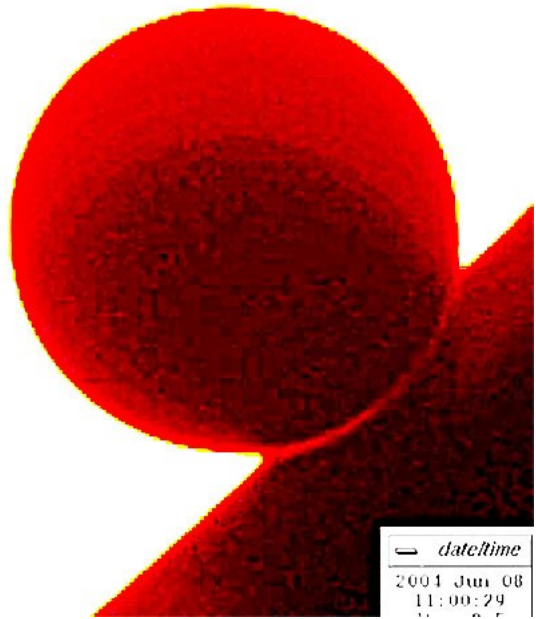

**Figure 4-** *Overexposed print of one of the images obtained with the TRACE mission by NASA during the 2004 transit, without performing any processing to show at the same time the disk and what is protruding outside the limb of about 1/10 of the diameter of the planet Venus as in the description of the arc by ML. See <12> and <13> for other qualitative images obtained by a posteriori computer compositing, with normally exposed parts of the solar disk adjoined by overexposed parts outside the disk, which hardly corresponds to what is observed visually without focal plane occultation.*

The 2012 transit provides even more recent images from the HMI experiment onboard the NASA SDO mission (SDO HMI Continuum | NASA, <10>). Images of the entire disk at 1" resolution were taken about every 5 min in the red continuum (617.3 nm). Focusing on the frames taken near contacts, one may examine the arc as it must have been visually apprehended in 1761: Figure 5 gives an example of (a) an original image and (b) the same after overexposure in order to see the arc.

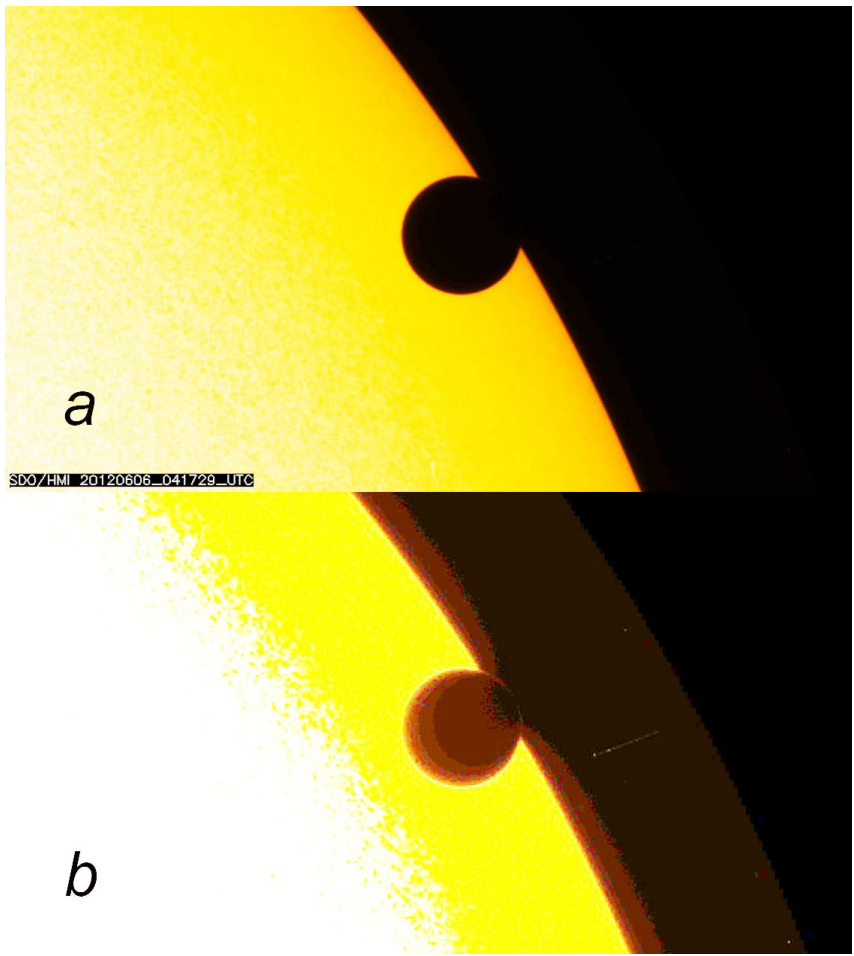

**Figure 5-** *Extract of an image from the HMI instrument of the SDO mission of NASA to show the arc during the egress of the planet Venus for the most recent transit of June 8, 2012. Images are taken with a narrow passband filter in the red. Colors are false colors in this presentation in order to show low contrast details. a/: with a reproduction of the original image to show the solar disk at the limb; b/: by overexposing the image to show the arc with an exit of the planet of about 1/10 of the value of its angular diameter, as for the relation of the phenomenon by ML in 1761. No processing is applied to enhance the fine details or improve the resolution. On image (b), one notices the stray light beyond the solar edge as well as on the image of the planet showing a pseudo « black- drop » effect on the stray light outside the disk, which is not due to a lack of angular resolution for fine details.*

### III- 4- The black drop parasitic effect

Inconsistencies in the various accounts of visual transit observations have been mentioned above, but no spurious effect has been as much in the news as the so-called black drop effect ( <7> <12> <14> for example). This effect is particularly noticeable just after the second contact and just before the third during the more "popular" and frequent transits of planet Mercury, as convincingly described by Pasachoff et al <12>. It has poisoned all Venus transit observations intending to accurately measure the instants of the contacts and by extension, measure the much sought-after Halley effect - or its variants based on parallax measurements on the planet Venus only. There is another and quite simple reason for this usurped popularity of the black-drop. Its origin is relatively well understood <12> and moreover, it is often observed and therefore well-known <7>. Mercury has no atmosphere and does not show a ML arc. On Figure 3 drawings or images 14 and 21 perfectly illustrate the effect noted during a Venus transit. The explanation is therefore to be found in image smearing, essentially due to image motion produced by turbulence or to diffraction and scattering in the optics and the Earth atmosphere. Imperfect focusing aggravates it, as may happen when using eyepieces equipped with an additional micrometric system at the prime focus, as was recommended for precise measurements of contacts as early as 1761, the very bright Sun adds blurring and dazzling effects. An edifying description of these associated parasitic effects can be found as early as 1762 in the scientific report of Chappe d'Auteroche, following his 1761 transit campaign in Tobolsk. Extensive studies by Pasachoff et al <12> and <14> have shown the black drop effect and its origin. Subtler seems to be a "sub-effect" as in Figure 3, drawing 15, which can be found in other authors, including in modern photographic observations. These are fleeting images captured just after the second contact and just before the third one. Even after the exit of the planet at the third contact (egress) it is still possible to detect these, but only on the stray light of the disk (figure 5b).

ML's observations never mentioned nor described the black drop effect, and described an excess of brightness, contrary to the black drop effect which corresponds to a "lack" of luminous flux in the image. On the basis of this important remark and in a description mentioning a bump on the solar edge or better a bright arc, it makes sense to interpret in as a refraction in a thick atmosphere, as first claimed by ML and others afterwards. Recent observations made with larger aperture telescopes do not report the "black-drop" effect when watching the transit of Venus (see further in part VI); it is reported only for the transit of the Mercury planet, much smaller in size, that incidentally does not have an atmosphere!

## IV- The atmosphere of Venus confirmed.

**IV-1- Confirmation after 1761**

After the transit of 1761 the existence of a Venusian atmosphere was confirmed independently, thanks to the observation of the planet during the lower conjunctions. Then, horns or even a luminous and thin shell or ring around Venus were at first attributed to the effects of refraction of solar rays, as for the ML arc. The first observations <7> are attributed to Johann Schröter (1745- 1816). He reported having observed horns on the Venus crescent beyond what could have been produced by the phase effect on the supposed ground of an illuminated planet without atmosphere, even if he was somewhat mistaken about the origin of the phenomenon. This phase effect is well-observed in the case of the Moon. H. N. Russell later <25> observed in 1898 the horns phenomenon with larger modern instruments. He is not immune from criticism on his interpretation in terms of scattering in the high layers of the Venus atmosphere, which led him to neglect refraction and to predict an atmosphere of Venus much more tenuous than the one known today, a widespread error at the time and even up to the time of the space missions in the 1970s.

The history of these observations of Venus' inferior conjunctions is somewhat epic and, fortunately, later on other famous observers repeated them, until today when they are a prime target for advanced amateurs. Figure 6, extracted from an atlas made several decades ago by P. Guérin (1967) <15>, perfectly shows a complete shell (envelope) during an inferior conjunction as well as the phase effect. This effect is very sensitive between the part located on the Sun side compared to the opposite part, located at only about 1 arc minute, i.e. 1/32 of the 32 arcmin solar diameter.

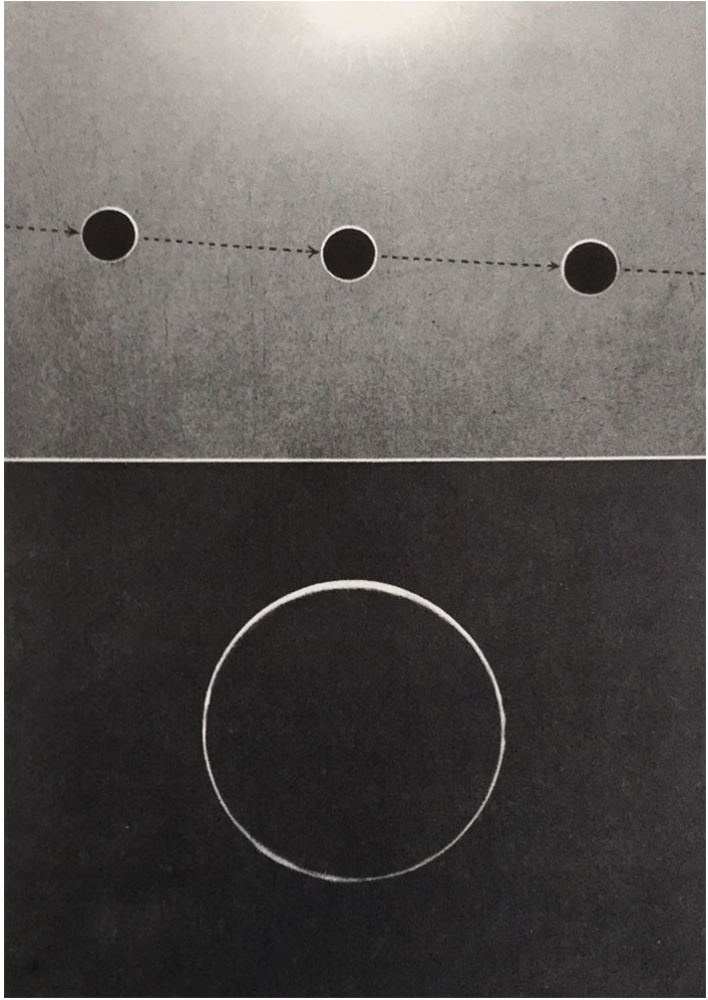

**Figure 6-** *Observation of the shell, or ring, or atmospheric aureole around the planet Venus during an inferior conjunction with the ring (aureole) almost complete, after L. Rudaux and G. de Vaucouleurs (1948) and P. Guérin (1967). Top: diagram showing the passage of the planet "under" the Sun (in the northern hemisphere) during the conjunction in projection on the sky. Bottom: image selected as close as possible to the Sun. Note the irregularities in the distribution of the light scattered in the atmosphere, as well as the beginning of the crescent partly due to the phase effect.*

This kind of observation, repeated at more recent conjunctions by astute and skillful observers with modest instruments, demonstrates unambiguously the importance of the thick atmosphere of Venus producing a ring of scattered light with a planet yet observed from "behind". Unique measurements of the polarization of the radiation from this shell were even made by B. Lyot as early as 1924 in his thesis work. These measurements, well summarized by A. Dollfus see <16>, indicate that the radiation was probably dominated by scattering on fine particles of about 2 micrometers in diameter.

**IV- 2- The Venus atmosphere today**

Contemporary data on the atmosphere of Venus (Figure 7) concur with the existence of such aerosol layer or cloud.

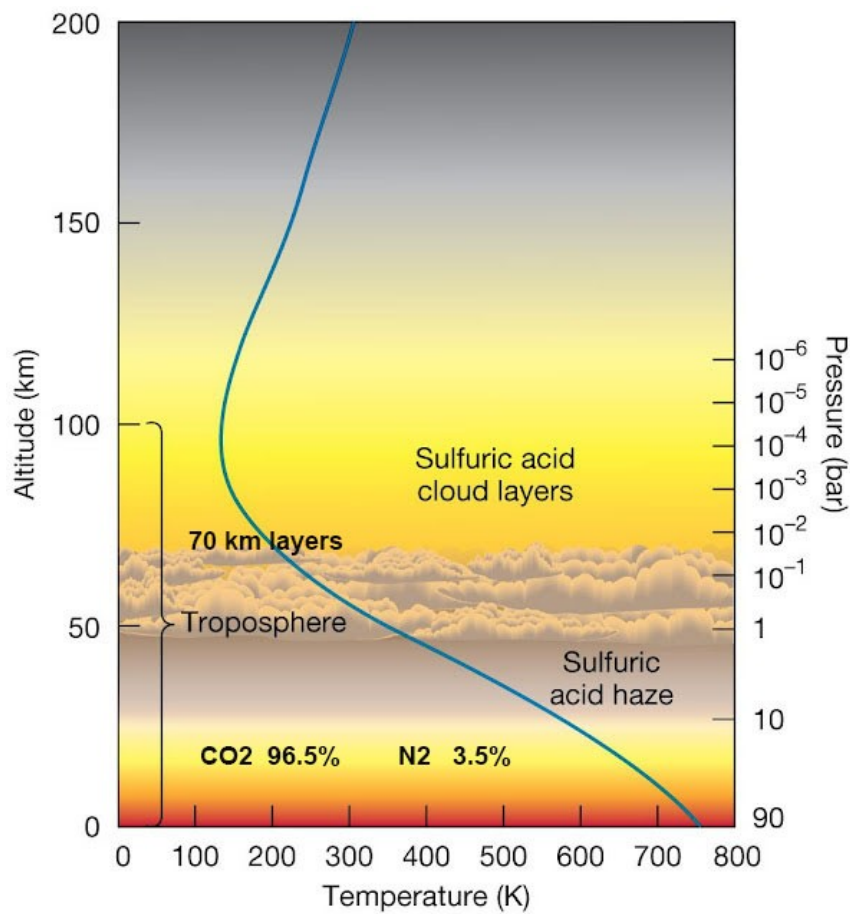

**Figure 7-** *Variations of composition, temperature and pressure in the atmosphere of Venus, between 0 and 200 km of altitude (from Pearson Education and T. Lombry- adaptation and added labels by the author). Notice the cloudy layer located immediately under the layers considered today as responsible for the arc of ML by refraction during the transit in front of the Sun (end of the 2$^{nd}$ contact at ingress and beginning of the 3$^{rd}$ contact at egress). These layers would be loaded with fine droplets of sulfuric acid giving a yellowish tint to the scattered light with a high albedo for the planet, close to 0.7.*

This layer is located above the thick lower atmosphere mainly composed of $CO_2$, between about 45 and 70 km altitude, It is moreover suggested that these particles would be clear (sulfuric acid droplets) or even white <10> <17>, located below an upper atmosphere that could extend up to 90 km and more. The problem of scattering by such particles is complex, even using a simplified Mie scattering approach as attempted by F. Link <1><2> at a time when the composition of the upper layers of the Venus atmosphere was very poorly known. Link made then an analogy with the case of the much fainter terrestrial atmosphere. To explain the ring or aureole of figures 4, 5, 6, including its thickness or intensity, it would be necessary today not only to imply the possible refraction in the lower layers of the atmosphere – the classical explanation introduced since ML and taken up 120 years later by Russell <<26> in the case of the transit - but possible effects of multiple scattering according to the phase and especially, the scattering in the higher layers by taking into account the aerosols, and the Rayleigh scattering (for a more complete discussion <5> <29>, <35>).

Finally let recall that Venus albedo is high, almost twice as high as the Earth albedo in the visible.

# V- On the modeling of the refraction phenomenon in the atmosphere of Venus and the important associated effects

In order to explain the observed effect of refraction near the solar limb, let now introduce more quantitative aspects.

**V- 1- The classical treatment: a transparent atmosphere**

The modeling of the refraction phenomenon in Venus atmosphere during a transit or passage of the planet has been classically tried (e.g. <1><2> <4> <6>). These efforts were partly inspired by analyzing phenomena like lunar eclipses in the case of the much fainter terrestrial atmosphere, and especially the passage of a bright star behind the atmosphere of a planet <23> <31>. An abundant literature exists, essentially using geometrical and optical descriptions, as it becomes clear by examining the extended studies described in Fr. Link's book <1><2> which, incidentally, takes up ML's suggestion. It seems difficult to add much to this classical work which includes distortion effects in 2D and a treatment of the phenomenon at a time when the specificities of Venus' atmosphere were not well known and, as already suggested by ML, this atmosphere could be denser than on Earth. Let observe that the classical interpretations are made with several simplifying hypotheses such as a fully transparent atmosphere for horizontal path, with a homogeneous and isothermal layer in order to simplify the geometrical parameters of the problem. These analyses ignore quantitative photometry and attenuation, by absorption and especially scattering, of the radiation passing through the various layers <5>.

Today, progress is possible thanks to the measurements made with radio waves which penetrate the cloudy layer and, above all, the space missions. These include the Soviet probes which progressively succeeded in penetrating the dense atmosphere of Venus and even reached the ground. There the conditions are extremely unfriendly for instruments beyond a few tens of minutes: the atmospheric pressure exceeds 90 bars and temperature reaches approximately 750K (figure 7). Without being exhaustive, let mention the most recent mission by ESA called "Venus Express" or VeX. It even carried a photometer (SPICAV/SOIR) allowing to probe the atmosphere by observing from a Venusian orbit the occultation of the Sun by the planet. These measurements could thus be used to interpret the last historical transit of Venus in front of the Sun of June 5-6, 2012, with a special mention to the thesis work of Ch. Pere <10> and the most recent article that summarizes his results see <21>, with an analysis that follows the main lines previously defined by the precursors such as P. Tanga and collaborators see <6>. Another interesting work in this context is the thesis of Marcq see <22> on the infrared spectra of the atmosphere of Venus confirming the important differences with the terrestrial atmosphere.

**V- 2- Some simple naive considerations to start with.**

As is sometimes done, reference to the case of the Earth's atmosphere seems useful in addressing the phenomenon when trying to evaluate the order of magnitudes. Both the diameter and the gravity are close for the sister planets. At ultimate grazing incidence (zenith angle is then 90°) it is well known that the light is refracted by typically 38' for an observer on the Earth at ground level <20> where the pressure is 1 bar, or, accordingly, by more than 1° for a ray tangentially crossing the whole atmosphere. In the limited case of the ML arc, we are dealing with a smaller amplitude effect <1><2>, that is produced by less dense layers than the ones at the Earth's ground, typically 100 times less, bending the light to give a "mirror" image[4] of the solar limb reflected above the planet limb of 60" diameter. This mirrored image appears above the cloud layers when the Venus atmosphere emerges from the Sun, as depicted in Figure 7. Assuming in first approximation that the pressure reflects the atmospheric densities, without paying attention to the temperature variations and to the composition, the relevant altitudes $h$ reaches 75 km, where the pressure is $10^{-2}$ bars, well above the cloud layer. Beyond Fig. 7 and making use of modern data on planetary atmospheres, the values taken from the classical and revised Allen's book <20> are quite appropriate. The pressures, temperatures, and densities of particles - nitrogen and oxygen molecules essentially - are given for the Earth's atmosphere in Table 11-21 of <20>, from the surface to 1000 km altitude. The review paper <24> can also be consulted. Let us consider the layer of the atmosphere that provides the radiation observed in the arc or the ring (or halo). Let $r$ be the radius of the planet reached by grazing rays but weakly bent by refraction; $\delta h$ the effective thickness of the layer concerned in a one-layer model, a value equal in first approximation, to the hydrostatic height scale at the considered altitude:

$$\delta h = \Pi T/\mu . g \qquad (1)$$

with $\Pi$ the universal constant of perfect gases, $T$ the absolute temperature of the layer, $g$ the gravity considered as constant in the thickness of the layer and $\mu$ the average molecular mass. The value of $\delta h$ is about 8 km for the Earth's atmosphere near the ground - they vary mostly with $T$, of course, but for refraction it is not very important. For the atmosphere of Venus, with a slightly lower gravity than the Earth and a somewhat different composition, with a dominance of carbon dioxide giving a higher $\mu$ because of $CO_2$ of higher molecular weight **44** compared to $N_2$ and $O_2$ for the Earth atmosphere, $\delta h$ varies between 4 and 7 km. To obtain $h$ one must take into account the effective length $d$ of the grazing rays. By simplifying and considering a straight trajectory in a homogeneous stratified curved atmosphere, $d$ is easily evaluated:

---

4   We use « mirrored » in the sense of « making an image ». It has nothing to do with the image reflected by a plan mirror by it could be compared to what is done with the grazing incidence concave mirrors used to produce an image or a spectrum in X- rays telescope.

$$d = (8 \cdot r \cdot \delta h)^{1/2} \quad (2)$$

This length d is derived with a slightly different approach giving a similar numerical coefficient (2π instead of 8) in <6> <10> <21> <31> and in the present special issue in the article by B. Sicardy <23>, formula (25).

To begin with, taking $\delta h$= 4 km for Venus at *T*# 200K and *r*= 6100 km, slightly above the cloud layer, *d* typically appears 100 times greater than $\delta h$. This could lead to another way to evaluate the height of the rays in the atmosphere of Venus, as discussed below.

**V- 3- An atmosphere with scattering and absorption**

Atmospheric attenuation was ignored above in the modeling of refraction effects during the transit of Venus; the atmosphere of Venus was considered transparent. This is a severe approximation to deal with a useful treatment of the refraction phenomenon observed after the light rays exit the atmosphere and certainly even more, to deal with the scattering that is superimposed on it. The case of the terrestrial atmosphere, well-illustrated by the Sun setting and rising, or the Moon sets and rises observed successively from the orbital space stations, e.g. < 25, 26, 27 >, is edifying in this respect. Furthermore, it seems important to fully address the issue for the future searches of extra-solar planets likely to show signs of life (exo-Earths). This will require the "deep" interpretation of light curves and spectra obtained during transits or passages of extra-solar planets in front of their parent star. For the advanced amateur, another application would be the interpretation of images and color index during total and even partial eclipses of the Moon.

To find the true height reached in the atmosphere by the emerging rays, it is necessary to account for the atmospheric attenuation at the time of the crossing, always neglected by the traditional treatment. This is indeed an unknown in the treatment of refraction and is often ignored by treating the outgoing radiation with an intensity coming from the source – here the surface of the Sun – *without any attenuation*. This is obviously false. The refraction of light rays depends, apart from a refraction coefficient to be considered later, on the density *N* of the transmitting gas. A compromise is therefore established between the maximum density encountered by the bent light rays, and the re-absorption and/or dispersion by light scattering, allowing the light rays pass. At the concerned densities, the effects which matter are:

i/ Molecular scattering (Rayleigh scattering, see for ex. <5> <20>) which is very sensitive in the violet and blue (efficiency in $\lambda^{-4}$ with $\lambda$ wavelength of the radiation). The directional pattern of the scattered intensities (or phase function) is in $(1+ \cos^2 \theta)$ [figure 8]. Hence and at a good approximation, the efficiency can be considered as isotropic over a range of a few degrees around the planet. Thus, the radiation scattered in the upper atmosphere produces a blue fringe of weak intensity, when compared to the refracted radiation as shown in <27> <28> for the case of the terrestrial atmosphere.

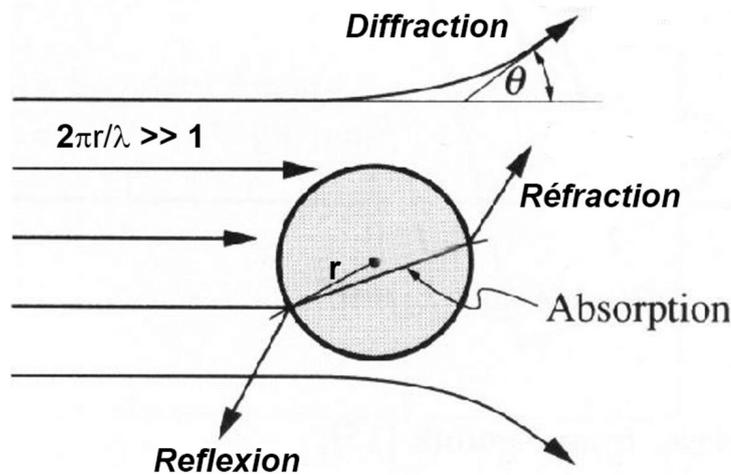

**Figure 8-** *Different processes identifiable during the interaction of the grazing solar rays in the upper atmosphere of Venus by meeting the droplets or aerosols, as present in the atmosphere given by figure 7. The refracted light is then much more dispersed than by the gas molecules producing the ML arc during the transits.*

ii/ Mie diffusion or scattering on aerosols, which for the Earth are of anthropic origin in addition to volcanic and meteorites origins (e.g. <29>), and on droplets or crystals of radii a> $\lambda$ (or even a>> $\lambda$). When these radii are quite comparable to the wavelength, the scattering efficiency is in $\lambda^n$ with n> 2, and the directional patterns give significant intensities forwards, as shown figure 8 and in Koutchmy & Koutchmy (1976) in <18>. In the case of Venus, these patterns must be convolved by the extent of the solar disk seen from the planet on about 45' with a non-negligible limb darkening. This component dominates at the time of the inferior conjunctions of Venus (figure 6).

In the case of the terrestrial atmosphere <20>, the optical thickness $\tau_0$ from ground at zenith, in the visible range, (see table 11.25 of <20>) is about 0.4 (i.e. a transmission of 0.64), which slowly improves towards the infra-red, except obviously in the molecular absorption bands of $CO_2$, abundantly present in the infra-red <22>. In the case of Venus, a focused study made by Ehrenreich et al (2011) <30> shows that the attenuation at the concerned altitudes (heights above 70 km) do not indicate a great difference compared to the terrestrial case, with Mie scattering largely dominating Rayleigh scattering above 500 nm wavelength. To simplify, we consider in equation (2) a linear relationship, in order to evaluate the attenuation along the path see and therefore treat the optically thin case. To obtain a typical transmission of 1/e, i.e. 4 times more than for a terrestrial type atmosphere, considering grazing on a homogeneous layer, it is enough to have a gas pressure – given by the density of the gas since we neglect the variation of temperatures – of 100/4 or 25 times less than the pressure of the sea leve. This corresponds to an altitude of about 20 km on Earth (table 11.21 of <20>). The corresponding pressure P is about 1/25 bars which gives typically 75 km of altitude for Venus, using figure 7. As indicated above, this altitude corresponds to the layers above the cloud layer found during surveys made with space missions, an agreement which is reassuring: the radiation will be able to go out of the crossed layer. Reference <30> gives a more extended treatment accounting for the wavelength, in particular towards the infra-red.

This evaluation of the order of magnitude can thus be refined to take into account the specific composition of the atmosphere of Venus and the obviously curved trajectory of the light rays, although, by definition, this curvature is limited according to the calculation of d (2). Nevertheless, further analysis shows that a multi-layer model should be developed <10>. Estimating orders of magnitude seems sufficient to evaluate the intensities (fluxes) observed in the arc and the variation of these intensities as a function of angular distance from the solar limb. Yet, observed intensities and the true width of the arc or aureole are indeed dominated by other factors, related to the considerable smearing showed by the point spread function of the instrument and including a defocusing factor, as well as by image motion due to the Earth's atmosphere. This factor, almost never analyzed, varies between values of the order of several arcsec for old observations and of the order of 0.2 or 0.3 arcsec for the best recent observations (transit of 2012) made with large and excellent telescopes on the ground and in space.

**V- 4- An improvement of the model.**

The study of the refraction of the rays starts from the discussion of the index $n_i$ of refraction of the atmosphere for the layer of density $N_i$ or better, of the so- called refractivity $(n_i-1)$ known to be related to $N_i$ :

$$(n_i-1) = (n_0 -1) \ N_i /N_0 \qquad (3)$$

where $(n_0 -1)$ and $N_0$ are the values taken at the reference level. The values are well known for the terrestrial atmosphere at sea level and above <20>; the refractive index is $n_0 = 1.000\ 29$ at $T= 0^0$ C, see <20>. For the Venus atmosphere with a composition where $CO_2$ dominates, $n_0$ is evaluated to be 1.000 45 at the reference level (1 bar). It is then interesting to evaluate the refraction angle that presumably should be bigger than, or of the same order as the amplitude of the emerging part of the planet Venus transiting before the Sun for the ML arc observation of a "mirrored" image of the solar limb. This amplitude was of the order of 1/10 of the Venus diameter or 6 arcsec. The refraction angle $\alpha_t$ (in radian) by definition < 1> <6> <31> etc., would be, making simplifying assumptions as a constant temperature $T_i$ of the effective atmospheric layer of length $d$ bending (refracting) the rays:

$$\alpha_t \ \# \ (n_i -1) \ d/ \delta h \qquad (4)$$

$d$ was defined above in (2) and its numerical value was found of order of 400 km. Let adopt a scale height $\delta h$, estimated equal to 4 km in the Venus atmosphere above the cloud layer at $T\#$ $200K$ at the altitude $h$ near 70 km (figure 7) ; this value is generally accepted in the literature <6> <10>). Using a typical density 3 orders of magnitude lower than the reference values, $(n_i -1) \ \# \ 10^{-3} (n_0 -1)$. This leads to a refraction angle of order of 10 arcsec, as needed to explain the ML arc. This value is still compatible with the evaluation[5] made by F. Link <1><2> for Venus, see his formula (5.17). Larger values for $\alpha_t$ would possibly better explain the sensitivity of the ML arc intensity, taking into account the large limb darkening effect at the extreme limb of the Sun especially in the blue region.

---

[5] Where the refraction angle is given in arcmin, the pressure in mm of Hg and the temperature in absolute unit. We take T= 190K, see figure 7, and get with his evaluated parameters for an atmosphere made of $CO_2$, and a pressure of 1 mm of Hg, a typical refraction angle of 17 arcsec.

Intensities of the arc observed outside the solar limb at distances much further out than the ML arc, see <6> <10>, drop quickly with the distance to the solar limb (see figure 9). A correct and more valuable calculation must be done with many layers, see <6> <10> and the Baum and Code paper <31>.

For a more elaborated treatment of the refraction effects in planetary atmospheres, with an application to the case of the ML arc and especially taking into account the differential refraction effect due to the radial gradient of atmospheric densities, the reader may refer to the article devoted to the analysis of refraction in astrophysics by Bruno Sicardy, in this special issue <23>. This elaborated theory is based on the analysis of the refraction for explaining the attenuation of light when observing the flux of a bright star at infinity crossing a transparent planetary atmosphere belonging to an object in the solar system. The light is then dispersed by the crossing of different layers of different gas densities and, accordingly, different values of refraction angles. There is an apparent decrease of the flux (magnitude) of the star when the beam approaches the surface of the planet, leading to the introduction of a parameter called the "half light level". It is applied to the case of a finite distance source of light like our Sun to explain the sort of "mirrored" highly distorted image of the limb producing an arc and further out, an aureola or ring effect, including the radial variations of the intensities of the refracted images see <6> <10> <21>. In case of the atmosphere of Venus a complication is appearing because one deals with a haze- polluted atmosphere with aerosols and clouds (figure 7 and 8 and section V-3 above).

Finally, to better understand the attenuation due to aerosols and show their importance, it is wise to consider again the phenomenon of halo that appears at lower conjunctions (Figure 6), to be compared with the arc of ML measured by Pere, Tanga et al (2011- 2016) see <6> <10> and figure 9. Very close to the solar limb, it is the refraction that produces an intense arc that can be compared to a sort of "mirrored" highly distorted image of the solar limb, while further away, with Venus well outside the disk, it is the scattering on aerosols that dominates in first approximation, making a bright ring appear around the planet. In fact, it would probably be necessary to consider the phase effects on the planet reflecting the solar light beyond the planetary limb and the double scattering, to better explain the intensities in the ring. Figure 9 summarizes different measurements, according to the distance to the limb of the Sun. Aureoles and rings have undoubtedly variable thicknesses, hence there is an obvious difficulty to compare the reported intensities. Ultimately, it would be necessary to correct the reported values for the considerable spread and blur in the images with respect to the true value of the thicknesses.

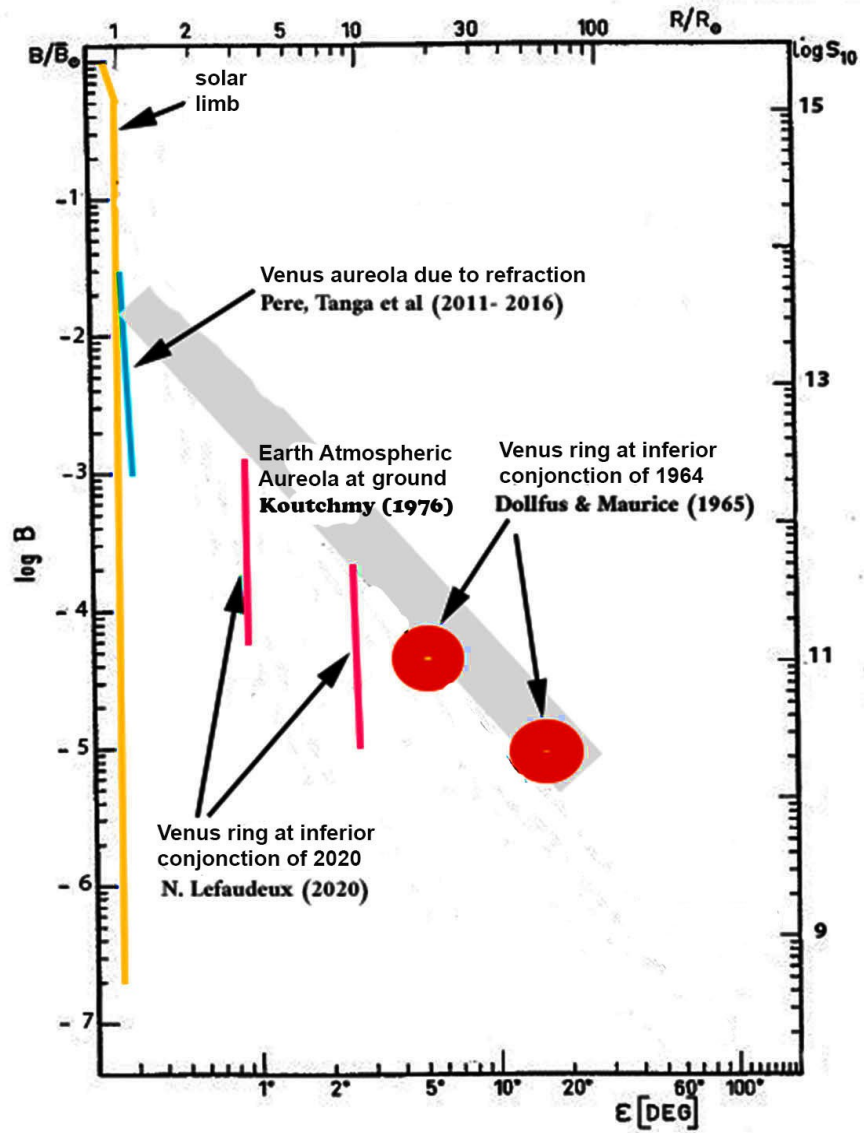

**Figure 9-** *A synthetic figure representing the radial variations of intensities (B-brightness or luminance of visible light) of the aureole and rings of Venus seen at inferior conjunctions, including transits at the limb of the sun. The arc of ML would be very close to the limb and it is therefore part of the radial variation shown in blue of the aureole according to Pere, Tanga et al (2011-2016) <6>. Further in light gray, typical variations of the luminance of the solar aureole with a clear daytime sky (according to Koutchmy 1976 <18>) at sea level or moderate altitude are shown to evaluate the contrasts of the rings of Venus during inferior conjunctions, according to Dollfus and Maurice (1965) <17> in red-brown and according to Lefaudeux 2020 in red see [2020 inferior conjunction of Venus - HDR astrophotography by Nicolas Lefaudeux (hdr-astrophotography.com)](#) . Note the log-log scale. On the right the luminance scale is in S10 units (number of stars of magnitude 10 per square degree) and at left, in units of the average brightness of the solar disk in the visible light. The position of the solar limb with its limb darkening is shown in yellow on the left.*

## VI- Modern measurements of the Lomonossov arc.

Let here conclude with two modern observations of the ML arc during the most recent transits of 2004 and 2012.

**VI- I- The ground-based high- resolution observation of 2004**

The Swedish Solar Telescope) (SST) is in 2004 the best ground-based solar telescope with an effective aperture diameter of 95 cm. It has taken many images from the excellent site of La Palma (Canary Islands), as well as filmed sequences, with many high frame rate images see [vt-2004.kva.astro.su.se/movies/](vt-2004.kva.astro.su.se/movies/). The wavelength (705 nm) is in the near infrared to reduce the smearing effects of the Earth's atmosphere). Figure 10 is an image extracted from the best film; it has been severely selected to eliminate images with motion and distortions, in order to provide the best possible quality at the time of the egress, the planet Venus appearing as described by ML, i.e. slightly off the limb of the Sun. No processing was made on this image. The "bump" or "blister" or "bulge" is easily guessed and, moreover, no trace of a black drop effect is perceptible. This is obviously due to the excellent quality of the image, since the smearing, although not perfectly uniform, is evaluated as an average 0.3'', but also to the fact that the planet is completely outside the solar limb. The ML arc is perfectly seen, and its intensity approximately corresponds to the intensity at the extreme solar limb, at 0"3 from the limb or a little less. At the wavelength of observation, the theoretical resolution of the telescope is 0.15" for the PSF full width at mid height. A measurement of the arc width gives 0.25'' which in 2004 provided the best evaluation ever made.

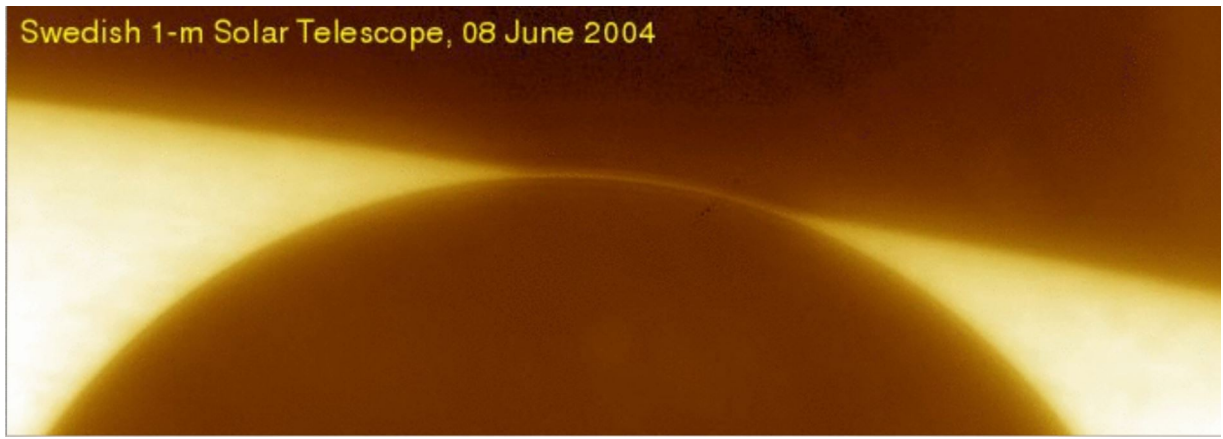

**Figure 10-** *Unretouched selected frame from the best film obtained on the SST of La Palma (Canary Islands) during the transit of 2004. Wavelength 705 nm. The arc of ML is possibly almost resolved with this beautiful image where nevertheless the quality is still limited by the image motion, the blur and the distortion (PSF equal to 0"3 on average whereas the theoretical resolution should be 0"15). The "bump" at the limb of the Sun is easily guessed. No black-drop effect perceptible even for a distance limb of Sun-outer limb of the Venus planet atmosphere of order of 1 arcsec and an intensity comparable to the solar intensity at the extreme limb.*

**VI-2- The space observation of 2012**

The best measurements so far was obtained in 2012 by the largest existing solar telescope in space, the SOT of the Japanese Hinode mission. Without any smearing due to the Earth's atmosphere, the spatial resolution is thus completely determined by the instrument alone and the results are perfectly reproducible, even at the shortest wavelengths of the narrow band imaging filter imaging where the resolution is less limited by diffraction.

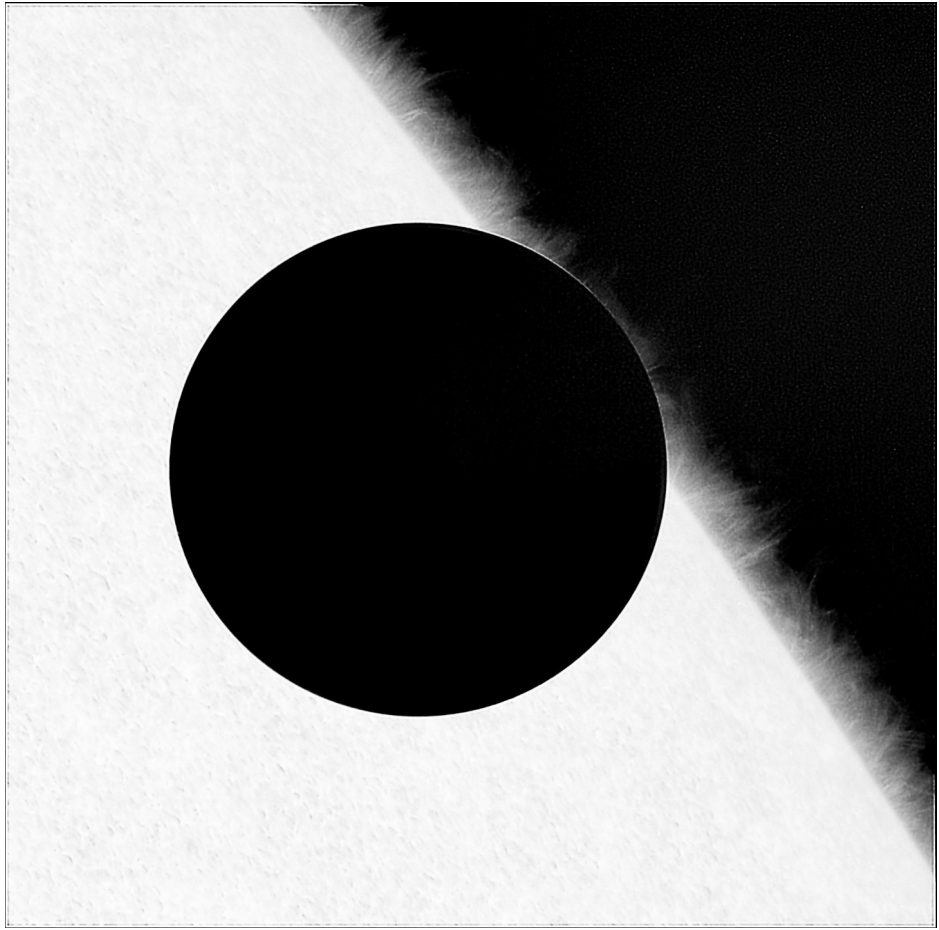

**Figure 11-** *Extract of one of the very high-resolution images obtained with the SOT of the Hinode mission, during the 2012 transit. The filter used (width 0.4 nm) is the one used to show the spicules or solar prominences thanks to the large emission in the broad chromospheric line called H of CaII. The chromospheric emission line fringe above the solar limb is on average 7 arcsec wide, i.e. about 4500 km on the Sun and about 3 times less at the distance of Venus. The ML arc is very thin and shows ~~perfectly~~ that the horns (parts of the arc closer to the solar limb) are thicker and more intense than the top of the loop or arc. Note that the solar disk is slightly overexposed on this reproduction in log I (logarithm of intensities), although the solar structures are still perceptible. No darkening at the extreme limb in this bandwidth and virtually no artifacts recorded in this image. Image by Goodarzy and Koutchmy <32> <33>.*

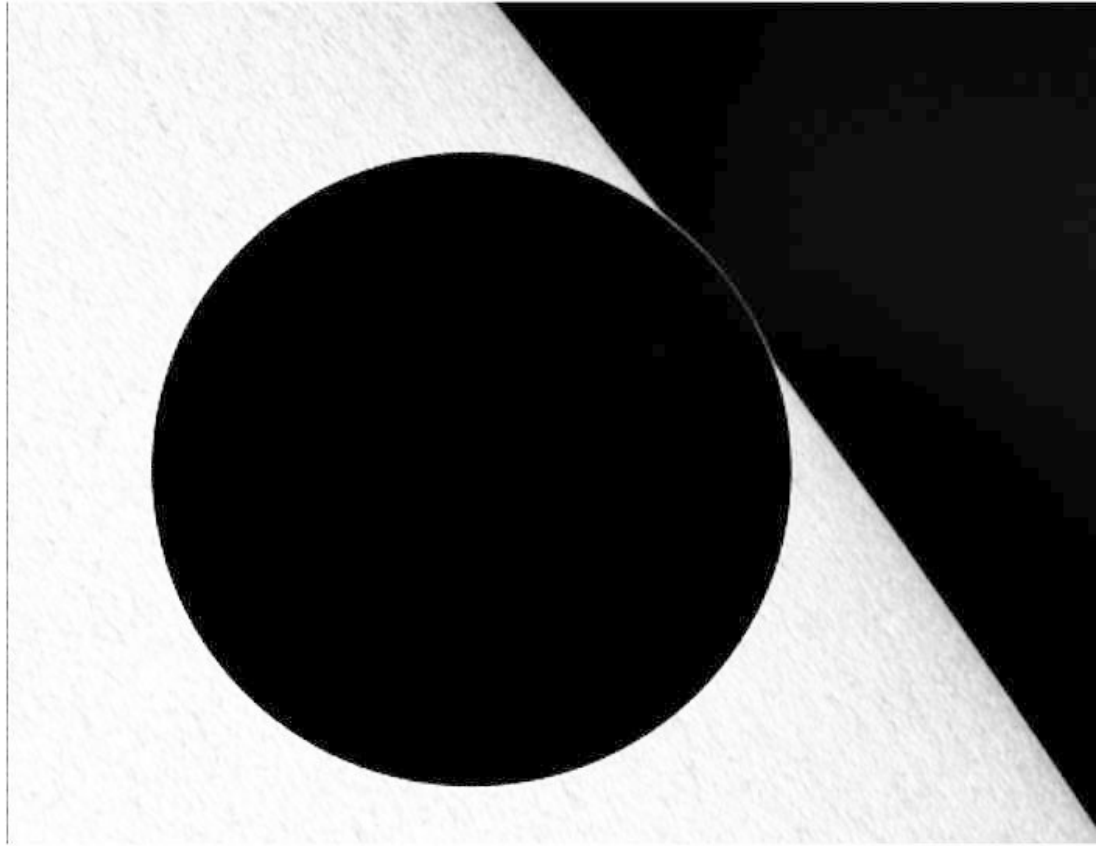

**Figure 12-** *Image of the planet Venus in the blue (450 nm; 0.4 nm FWHM filter) of an image obtained with the SOT of the Hinode mission after the 3d contact at egress. Illustration to show the effects on the solar disk at the limb with a large center- limb darkening effect reaching a large value at the extreme limb in the arcsec range. The refraction effects in the atmosphere of the planet is shown by selecting a good exposure and a low contrast to show the variations all around the planet at different distance of the solar limb. Note the elongated granulation near the solar limb due to the limb shortening effect on granules.*

Using these data, an extensive study was done in the thesis of Hadis Goodarzi <32> <33> about umbral dots in sunspots. The PSF[6] has been deduced <32> during the transit of Venus, for different filters, with two assumptions:, i/ that no stray light should appear on the disk of Venus in projection on the solar disk nor on the sky background beyond the solar limb; ii/ using a very narrow solar limb as given by the most recent solar atmosphere model, in particular in the vicinity of the so-called temperature minimum region where the hypothesis of a stratified atmosphere in hydrostatic equilibrium is still perfectly valid. This region is located at about 500 km of altitude in the Sun with respect to the reference model level at the disk center. The author deduces a real solar limb smearing, similar to what the most recent solar atmosphere model gives, found to be 0"215 as seen from Earth. The PSF has been deduced with an excellent signal-to-noise ratio and can therefore be applied to the images, in order to improve the resolution by 2D deconvolution see <33>. The resulting images have a uniform resolution corresponding to the theoretical resolution given by the width at half intensity of the PSF, accounting also for the far wings of the smearing function due to diffraction effects on the mirrors, the secondary's spider supports and the filter, up to a significant distance quantified in arc minutes for intensity levels below 1/1000 see <32>.

---

6   The point spread function or PSF was deduced without assuming a gaussian type behavior in its far wings that extend to large distances, but assuming a shift- invariant and axi- symmetric PSF with a core corresponding to a diffraction limited resolution for the correspondang wavelength.

To discuss the ML arc with the best possible resolution, we selected images obtained with the blue filter in the pseudo continuum - avoiding as much possible the deep dark lines of the Fraunhofer spectrum - of the solar spectrum at 460 nm with a bandwidth of 0.4 nm. The deconvolution of the images was done by seeking the best compromise in terms of signal to noise ratio and avoiding the Gibbs effects[7]. The result is interesting (figure 12 illustrates the effects). The ring or arc or halo is observed not only outside the solar edge, but also inside the disk but with a limited extent. The interpretation seems obvious as long as one takes into account the center-limb effect of the solar disk, quite strong in the blue: the refraction reproduces a portion of the solar disk farther from the edge and therefore brighter as predicted by Link see <1><2>. The parts farther from the limb hardly show this effect anymore and in fact the solar granulation is visible up to the edge of the Venus image as for example at the side opposite to the solar edge. Thus, the specificities of the refraction mechanism in the atmosphere of Venus are illustrated on this figure. Let us recall that the spatial resolution in these images, of the order of 0.1", is largely superior to all that could be reported with visual observations of the transit, by about a ratio 30, which makes it a little illusory to make quantitative comparisons with former observations, without saying that these new results are photometrically (quantitatively) correct see figures 13 and 14. It should be noted that the refraction-bent solar rays seen in the arc and partially on the disk are hardly attenuated in the

---

[7] See the definition in Gibbs phenomenon - Wikipedia

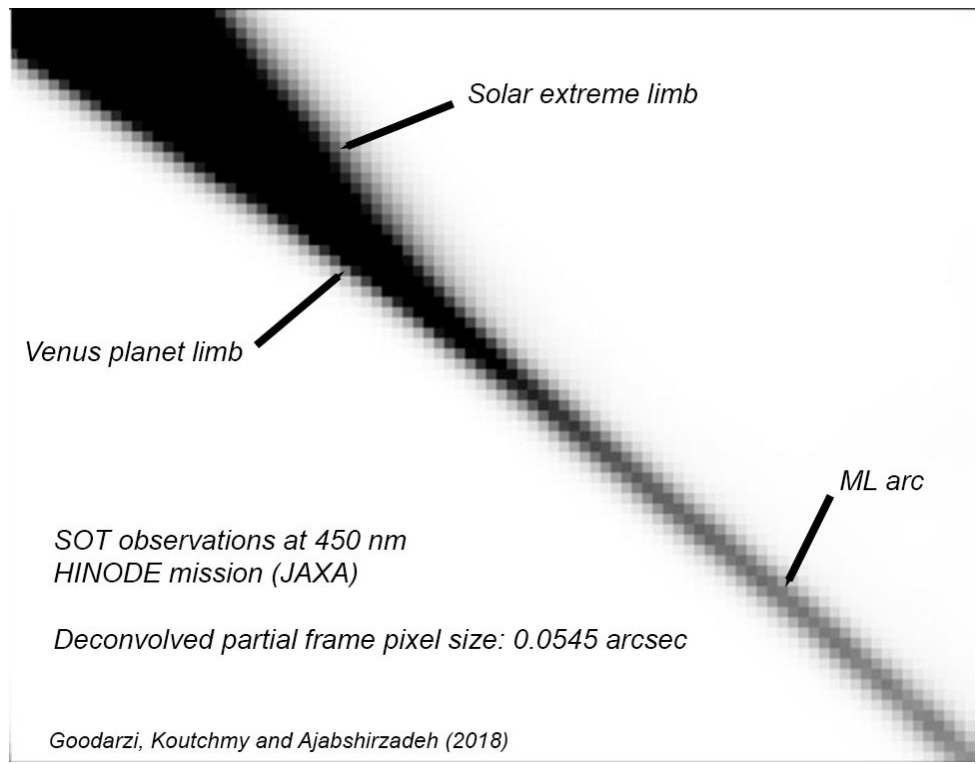

**Figure 13-** *16 times enlarged part of figure 12 in negative with high contrast to show the arc of light due to refraction, about the solar disk near the limb and outside the disk (ML arc) after deconvolution. The scale is given by the size of the pixel, i.e. 0"0545 or 11 km at the planet Venus distance, permitting a quick evaluation of the apparent width of the arc.*

Venus atmosphere. They come from high layers where the pressure is low and refraction is limited as in <6> and <10>. Without extinction, the intensity of bended rays giving the arc would show the solar intensity.

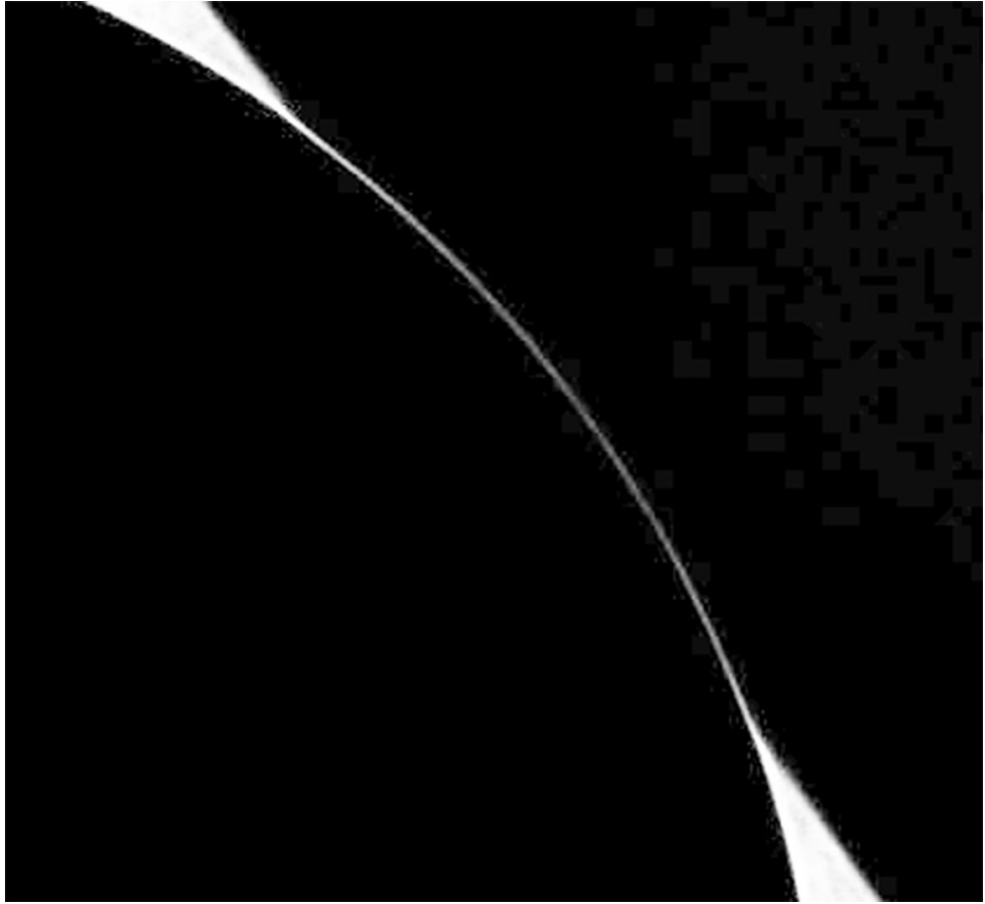

**Figure 14-** *Partial frame of the deconvolved image of figure 12 in positive to show the whole ML arc outside the solar disk. Note the slightly heterogeneous distribution of light along the arc suggesting that the resulting resolution after deconvolution is barely sufficient to guess the details of the heterogeneous atmosphere of Venus at the altitude where the refraction is efficient. The relative intensities recorded in the arc vary in relative units from 370 (solar side) to 80 in the arc near its top; at the center of the Sun the average disk intensity is 800. On the disk of Venus the intensities are almost zero, as in the sky beyond the solar limb, after deconvolution. The impression here is that the arc outside the solar limb is attenuated in a significant way, contrary to the case of figures 10 made in the near IR or of figures 11 made in the broad chromospheric line of CaII H with a much less limb darkening of the solar disk.*

To go further and derive some quantitative values on the ML arc, one may enlarge considerably the images, about 16 times, showing the pixels. Figure 13 shows a small part of figure 12 located on the arc, both outside the solar disk and on the disk, upper part, with the same orientation. The scale is now given by the pixel size equal to 0"0545 or 11.1 km on Venus. We estimate the half width at mid-intensity of the arc outside the disk at about 2.5 px or 28 km. In fact, this width varies along the arc around 2.5 px and it would be necessary to make an analysis all along the arc to determine these inequalities or dispersion. They presumably reflect effects of the cloud cover, which is more or less changing with the retrograde rotation of the atmosphere in 4 days of the atmosphere, as well as the position in latitude and longitude. In fact, other diagnostics can be proposed such as obtaining spectra of different regions of the atmosphere in projection on the disk <30> or on the arc of ML and thus outside the solar disk. This has been attempted with the Yunnan NVST ground-based 98 cm aperture telescope, see Bazin et al. (2016) <34>.

These spectroscopic results are numerous, but their analysis is not yet finished, and, after cleaning of instrumental effects, they will be published later.

## VII- Conclusions

The study of the phenomena occurring during the transits of Venus in front of the Sun is relevant to the contemporary research in exoplanets, including exo-Earths and exo-Solar systems, when transits are observed e.g. <35>. This theme becomes a priority in astronomical and astrophysical research for the coming decades, on the ground and in space. Thus, the discovery of a rather thick atmosphere around Venus in 1761 by ML was a major step on the rocky path which may someday lead to another invaluable discovery long awaited, one that would ultimately provide evidence of life elsewhere than on our planet. This is an exhilarating and certainly historic task that Academician M. Lomonosov has accordingly contributed to tackle by making the first step: suggesting the existence of an atmosphere around our sister planet Venus.

In this sense, as Lev Zeliony <36> has recently pointed out, ML can be considered a precursor of the philosophical current known as the Russian cosmism - opposed to Marxist thought - at the origin of the space exploits of the former USSR, which the Russian Federation of today is trying to take over. As mentioned in <36>, not only has this led to the huge investments made in the past with the "Venera" program, but the IKI (Moscow Center for Space Studies) is preparing two new missions at the end of the present decade to the planet Venus for searching there signs of life, in its atmosphere and even on the ground, in response to the hopes of habitability largely but somehow exaggeratedly expressed in the 18$^{th}$ century by ML and his contemporaries. Note that NASA has also planned a new space mission to Venus after 2030, with the goal of landing a probe on the ground. It is also speculated that the planet may be showing much sought-after signs of microbial life even in the very hostile atmosphere at the surface <36>.

In this conclusion, let also remember in passing that Halley's method, so much claimed and developed on the transits of Venus to measure absolute distances in the solar system, turned out to be based on a kind of misunderstanding for the precise measurements and dating of contacts. The fault was to neglect the effects due to the thick atmosphere of the planet and the effects of refraction, quite unequal otherwise, effects which had been neither anticipated nor even definitively recognized for more than one century. Let us finally notice that what we systematically described in this paper as the ML arc is based on modern knowledge and measurements. It cannot be straightforwardly identified with what Lomonossov described in 1761 claiming that the planet Venus owns an atmosphere as the Earth, or even thicker. Although the method he used and his instrumentation were optimum for his time, what really Lomonossov could have seen with his eyes and brain remains partly unproven and possibly affected by a kind of "psycho- physiological" effect not considered in this article and likely to be of little interest.

**Acknowledgements**

Many people helped me during the elaboration of this article. First of all, I thank my University teachers of the 1960ies and among them, the late F. Link who introduced me to the effects of refraction for the Earth's atmosphere and showed me the «vase atmosphérique» concept when rising together to the Pic du Midi Observatory at sun rise. I thank William Sheehan for meaningful discussions that sometimes went far beyond Venus. Remarks of Jay Pasachoff were also greatly appreciated. Hadis Goodarzi, Paolo Tanga, François Sèvre and Nicolas Lefaudeux provided material, advices and discussions for preparing the paper; Bruno Sicardy has been a meaningful and professional reviewer of the first draft of this paper. David Valls-Gabaud and Cyrille Bazin worked on the topic together with the author and the Chinese team headed by Zhi Xi, to perform the Yunnan experiment. Finally, Pierre Léna and Luc Dettwiller friendly helped me with the editing.

# References

1 F. Link (1969) Eclipse Phenomena in Astronomy, Ed. Springer- Verlag, **5-6** pp. 215- 216

2 F. Link (1959) "Transits of Venus on the Sun", BAIC, **10**, 4, pp. 105- 115

3 ESO. www.eso.org/public/outreach/eduoff/vt-2004/

4 V. Shiltsev (2014) "The 1761 Discovery of Venus' Atmosphere: Lomonosov and others", *Journal of Astronomical History and Heritage*, **17(1)**, 85-112 (2014).

5 W.B. Hubbard et al (2001) Theory of Extra-solar Giant Planet Transits", Astrophys. J. **560**, 1, 413-419

6 P. Tanga_, T. Widemann, B. Sicardy, J. M. Pasachoff, J. Arnaud, L.Comolli, A. Rondi, S. Rondi, P. Sutterling (2012) Sunlight refraction in the mesosphere of Venus during the transit on June 8th, 2004, Icarus **218**, 1, 207-219

7 Luce Langevin, Lomonossov, 1711-1765. Sa vie, son œuvre, Paris, Éditions sociales, March 1967 (1$^{st}$ ed. 1967), 320 p.

8 William Sheehan and John Westfall (2004) "The Transits of Venus", Prometheus book Publ.

9 Mikhail V. Lomonosov (1761) "The appearance of Venus on Sun as it was observed at the St. Petersburg Emperor's Academy of Sciences on May 26, 1761 (St. Petersburg, Russia). Petersburg: Academy of Science Office)". In Lomonosov, M.V., 1955. *Memoirs in Physics, Astronomy and Instruments Building*. Leningrad, U.S.S.R. Academy of Sciences (see also references in <3 >)

10 Christophe Pere (2016) "Study of the atmosphere of Venus using a refraction model during the transit in front of the Sun of June 5-6, 2012" Thesis of the University of Nice Sophia Antipolis

11 Charles Lagrange (1884) "Transit of Venus on December 6, 1882. Documents and observations. Mission of Chile." Ann. Obs. de Bruxelles, 5b, pp 81- 123
https://ui.adsabs.harvard.edu/#abs/1884AnOBN...5b..81N/abstract

12 Jay M. Pasachoff, Glenn Schneider, and Leon Golub (2004) "The black-drop effect explained" in *Transits of Venus: New Views of the Solar System and Galaxy Proceedings IAU Colloquium No. **196**, 2004 D.W. Kurtz, ed.*

13 J.M. Pasachoff, G. Schneider & Th. Widemann (2011) "High Resolution Satellite imaging of the 2004 Transit of Venus and asymmetries in the Cytheran Atmosphere" , The Astronomical Journal, **141:112** (9pp), 2011 April doi:10.1088/0004-6256/141/4/112

14 Jay M. Pasachoff, William Sheehan (2012) "Lomonosov, the discovery of Venus's atmosphere, and Eighteenth century transits of Venus", *Journal of Astronomical History and Heritage*, 15(1), 3-14 (2012).

15 Pierre Guérin (1967) "Planètes et Satellites", *under the direction of*, Larousse Paris Ed. p. 158

16 Audouin Dollfus (1951) "Polarization studies of Planets" Chap. 9, 343- 399 in Planets and Satellites, edited by Gerard P. Kuiper and Barbara M. Middlehurst, Chicago: The University of Chicago Press, 1961

17 A. Dollfus and E. Maurice (1965), "Physique planétaire- Etude de l'allongement des cornes du croissant de Vénus en juin 1964", CRAS, Paris, t. **260**, 3, 427

18  O. Koutchmy and S. Koutchmy (1976) "The coronal aureola in the time of total solar eclipse" Astron. Astrophys. Suppl. **13**, 295- 303

19 Aurélien Stolzenbach (2016) "Etude de la photochimie de Vénus à l'aide d'un modèle de circulation générale", Planétologie.   UPMC - Université Paris 6 Pierre et Marie Curie, Thesis 2016, in French.  See also https://www.researchgate.net/profile/ Aurelien Stolzenbach

20 G. Schubert and R.L. Walterscheid (2000) "Earth", Chap. 11 in Allen's Astrophysical Quantities, Arthur N. Cox Ed. pp. 239- 292, AIP Press and Springer

21 C. Pere, P. Tanga, Th. Widemann et al. (2016) "Multilayer modeling of the aureole photometry during the Venus transit" Astron. Astrophys. 28528

22 E. Marcq (2006) " Prélude à la mission Vénus Express : étude de l'atmosphère par spectro- imagerie infrarouge " Thèse de doctorat Université Paris VII, LESIA Obs. de Paris- Meudon.

23 B. Sicardy (2022) "Study of atmospheres in the solar system, from stellar occultation or planetary transit", Comptes-Rendus Physique, XXXXXX

24 A.T. Basilevsky, J.W. Head (2003). "The surface of Venus"  Rep. Prog. Phys. **66** (10): 1699-1734.

25 H.N. Russell (1899) " The Atmosphere of Venus ", *Ap.J.* 9, 284- 299

26 S. Koutchmy and G.M. Nikolsky (1983) "The night sky from Salyut 7", Sky & Telesc. 1983, vol. 65, no. 1, pp. 23-25

27 G.M. Nikol'skii, Yu. V. Platov, M. Belmahdi, V.V. Butov, E.S. Vanyarkha, E.S. Dzhanibekov, J-L. Chretien and S. Koutchmy (1987) "Stratospheric luminescence observed on the Salyut-7 orbital station", Issled. Zemli Kosmosa, 1987, no. 6, pp. 3

28 Yu. V. Platov, S. Kouchmy and S. Sh. Nikolayshvili (2017) "Atmospheric Emission Layers according to Photographic Observations from the International Space Station", Geomagnetism and Aeronomy, 2019, Vol. **59**, No. 3, pp. 351-355

29 Lei Li, Huizhang Che, Xindan Zhang et al, 16 authors (2021), "A satellite-measured view of aerosol component content and optical property in a haze-polluted case over North China Plain," Atmospher. Research, **266**:105958

30 D. Ehrenreich, A. Vidal-Madjar, T. Widemann et al. (2012) "Transmission spectrum of Venus as a transiting exoplanet" Astron. Astrophys. **537**, L2- 8

31 W.A. Baum and A.D. Code (1953) "A photometric observation of the occultation of sigma Arietis by Jupiter", ApJ **58**, 1208, pp. 112- 116

32 H. Goodarzi, S. Koutchmy, A. Adlabshirizadeh (2015), "Improved SOT (Honode mission) high resolution imaging observations", Ap&SS. **358**, in arXiv 1290160

33 H. Goodarzi, S. Koutchmy and A. Adjabshirizadeh (2018), "Proper motions of Sunspots Umbral Dots at High Temporal and Spatial Resolution", ApJ. Vol. **860**, Issue 2, article id. 168, 9 pp arXiv: 1807.05531

34 C. Bazin, X. Zhi, D. Valls-Gabaud, S. Koutchmy, P. Rocher, Z.Y. Zin, Y. Fu, L. Yang,G.Q. Liu, Z. Liu, K. Ji and H. Goodarzi , 2014 "The June 6 2012 Transit of Venus : Imaging and spectroscopic Analysis of the upper Atmosphere Emission " in SF2A 2014 J. Ballet, F. Bournaud, F. Martins, R. Monier and C. Reyle (eds)

35 Yan Bétrémieux and Lisa Kaltenegger (2014) "Impact of Atmospheric Refraction: how deeply can we probe Exo- Earth's Atmospheres during primary Eclipse Observations?"   The Astrophysical Journal, **791**:7 (12pp), 2014 August 10

36 Lev Zeliony (2021) « "At the edge of the Sun, a pupyr appeared.", the Kommersant, 28 nov. 2021 in Science,  Lomonosov Readings, held in Arkhangelsk and Kholmogory.